\documentclass[twocolumn,showpacs,preprintnumbers,amsmath,amssymb,superscriptaddress,prb]{revtex4}

\usepackage{graphicx,dcolumn,bm}

\newcommand{\nl}{\nonumber \\}
\newcommand{\be}{\begin{equation}}
\newcommand{\ee}{\end{equation}}
\newcommand{\bea}{\begin{eqnarray}}
\newcommand{\eea}{\end{eqnarray}}
\newcommand{\bsube}{\begin{subequations}}
\newcommand{\esube}{\end{subequations}}
\newcommand{\Fig}[1]{Fig.\,\ref{#1}}
\newcommand{\Eq}[1]{Eq.\,(\ref{#1})}
\newcommand{\Eqs}[1]{Eqs.\,(\ref{#1})}
\newcommand{\alf}{\alpha}
\newcommand{\sgm}{\sigma}
\newcommand{\omg}{\omega}
\newcommand{\Omg}{\Omega}
\newcommand{\Gam}{\Gamma}
\newcommand{\dlt}{\delta}

\newcommand{\Lmd}{\Lambda}
\newcommand{\lmd}{\lambda}

\newcommand{\vrho}{\varrho}
\newcommand{\vpl}{\varepsilon}
\newcommand{\epl}{\epsilon}
\newcommand{\la}{\langle}
\newcommand{\ra}{\rangle}
\newcommand{\nL}{{n_{\rm L}}}
\newcommand{\nR}{{n_{\rm R}}}
\newcommand{\rmi}{{\rm i}}

\newcommand{\rmL}{{\rm L}}
\newcommand{\rmR}{{\rm R}}

\newcommand{\rmB}{{\rm B}}

\newcommand{\etaL}{\eta_{\rm L}}
\newcommand{\etaR}{\eta_{\rm R}}
\newcommand{\GamL}{\Gamma_{\rm L}}
\newcommand{\GamR}{\Gamma_{\rm R}}

\begin{document}

\title{Renormalized dynamics in charge qubit measurements by a single
electron transistor}

\author{JunYan Luo}
\email{jyluo@zust.edu.cn}
\affiliation{School of Science, Zhejiang University of Science
  and Technology, Hangzhou 310023, China}
\author{HuJun Jiao}
\affiliation{Department of Physics, Shanxi University, Taiyuan, Shanxi 030006, China}
\author{Jianzhong Wang}
\affiliation{School of Science, Zhejiang University of Science
  and Technology, Hangzhou 310023, China}
\author{Yu Shen}
\affiliation{School of Science, Zhejiang University of Science
  and Technology, Hangzhou 310023, China}
\author{Xiao-Ling He}
\affiliation{School of Science, Zhejiang University of Science
  and Technology, Hangzhou 310023, China}

 \date{\today}

 \begin{abstract}
 We investigate charge qubit measurements using a
 single electron transistor, with focus on the
 backaction-induced renormalization of qubit
 parameters.
 It is revealed the renormalized dynamics leads to
 a number of intriguing features in the detector's
 noise spectra,  and therefore needs to be accounted
 for to  properly understand the measurement result.
 Noticeably, the level renormalization gives rise to
 a strongly enhanced signal--to--noise ratio, which
 can even exceed the universal upper bound imposed
 quantum mechanically on linear-response detectors.
 \end{abstract}


\maketitle

\section{\label{thsec1}Introduction}

 The issue of measurement lies at the heart of the
 interpretation of quantum mechanics.
 The recent upsurge in the interest to the quantum
 computation has attracted renewed attention to the
 problem of quantum measurement.\cite{Los98120,Kan98133}
 Various schemes have been proposed for fast readout
 of a two--level quantum state (qubit).
 Among them, especially interesting are electrometers
 whose conductance depends on the charge states of a
 nearby qubit, such as quantum point contacts
 (QPC)\cite{Ale973740,Gur9715215,Buk98871,Goa01125326,%
 Ave05126803,Pil02200401,Cle03165324,Li05066803,Luo09385801}
 and single electron transistors
 (SET).\cite{Shn9815400,Mak01357,Cle02176804,%
 Jia07155333,Gil06116806,Gur05073303,%
 Jia09075320,Oxt06045328,Ye10050726}
 It has been shown that the SET detector is better
 than QPC in many respects,\cite{Dev001039} and has
 already been used for quantum measurements.\cite{Sch981238}

 So far, theoretical description of the SET detector has been
 mainly focused on the backaction--induced dephasing and
 relaxation, which, from the perspective of information, are
 consequences of information acquisition by
 measurement.\cite{Gil06116806,Gur05073303,Jia09075320}
 Actually, there is another important backaction which
 renormalizes the internal structure of the
 qubit,\cite{Luo09385801} and is often disregarded in the
 literature.
 However, this renormalization effect is of essential importance,
 since it can crucially influence the dynamical process of
 quantum measurement.
 It is, therefore, required to have this feature being properly
 accounted for in order to correctly  understand and analyze the
 measurement results.

 In this context, we examine the renormalized dynamics of
 qubit measurements using an SET detector.
 The intriguing dynamics arising from the renormalization
 is manifested unambiguously in the noise spectral of
 the detector output.
 It is demonstrated that in the low bias regime, the noise
 peak reflecting qubit oscillations shifts markedly  with
 the measurement voltages.
 Furthermore, a peak at zero--frequency arises, as the level
 renormalization results in a so--called quantum Zeno effect.
 The output noise spectral allows us to evaluate the
 ``signal--to--noise'' ratio, which  provides the measurement
 of detector effectiveness.
 Noticeably, it is revealed that for the SET detector, the
 level renormalization leads to a considerably enhanced
 effectiveness, which can even exceed the upper bound imposed
 on any linear--response detectors.\cite{Kor01165310}

 The Letter is structured as follows. The measurement setup
 and model Hamiltonian are introduced in the next Section. We
 sketch the quantum master equation approach in Section \ref{thsec3}.
 The results and discussions are presented in Section
 \ref{thsec4}, which is then followed by the summary
 in Section \ref{thsec5}.

 \section{\label{thsec2}Model Description}

 \begin{figure}
 \includegraphics*[scale=0.4]{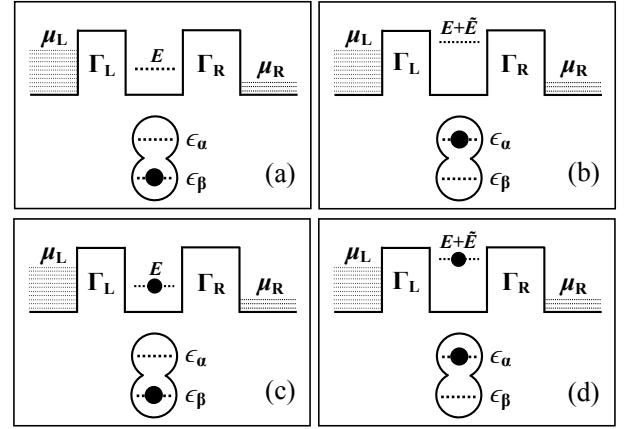}
 \caption{\label{Fig1}Schematic setup for a solid--state charge
 qubit measurements by an SET detector.
 Possible electron configurations of the measured qubit and
 SET dot are shown in (a)--(d), respectively.}
 \end{figure}

 The setup for the measurement of a charge qubit (an electron in a
 pair of coupled quantum dots) by a single  electron transistor is
 schematically shown in \Fig{Fig1}. The entire  system Hamiltonian
 reads $H=H_{\rm S}+H_{\rm B}+H'$. The first component
 \bea
 H_{\rm S}=\frac{1}{2}\epl\,\sgm_z+\Omg\,\sgm_x
 +(E+\tilde{E}|\alf\ra\la\alf|) d^\dag d
 \eea
 models the qubit, SET dot, and their coupling, with
 pseudo--spin operators
 $\sgm_z\equiv|\alf\ra\la\alf|-|\beta\ra\la\beta|$ and
 $\sgm_x\equiv|\alf\ra\la\beta|+|\beta\ra\la\alf|$.
 For the qubit, it is assumed that each dot has only
 one bound state, i.e., the logic states $|\alf\ra$
 and $|\beta\ra$,  with
 level detuning $\epl$ and interdot coupling $\Omg$.
 The SET works in the strong Coulomb blockade regime,
 and only one level is involved in the transport.
 Here, $d$ ($d^\dag$) is the annihilation (creation)
 operator for an electron in the SET dot.
 The single level (or equivalently, the transport
 current) depends explicitly on the qubit state,
 as shown in \Fig{Fig1}.
 It is right this  mechanism that makes it possible
 to acquire the  qubit--state information from the
 SET output.

 The second component $H_{\rm B}=\sum_{\ell=\rmL,\rmR}\sum_k
 \vpl_{\ell k}c^\dag_{\ell k}c_{\ell k}$ depicts the left and
 right SET electrodes.
 Here $c_{\ell k}$ ($c^\dag_{\ell k}$) denotes the annihilation
 (creation) operator for an electron in the electrode
 $\ell\in\{\rmL,\rmR\}$.
 The electron reservoirs are characterized by the Fermi
 distribution $f_{\rmL/\rmR}(\omg)$.
 We set
 $\mu^{\rm eq}_\rmL=\mu^{\rm eq}_\rmR=0$ for the equilibrium
 chemical potentials.
 An applied measurement voltage $V$ is modeled by different
 chemical potentials in the left and right
 electrodes $\mu_{\rmL/\rmR}=\pm V/2$.

 Electrons  tunneling  between SET dot and electrodes is described
 by the last component
 $H'=\sum_{\ell k} (t_{\ell k}c^\dag_{\ell k}d+{\rm h.c.}) \equiv
 \sum_\ell (F_\ell^\dag d+{\rm h.c})$, where $F_\ell$ and $F_\ell^\dag$
 are defined implicitly.
 The tunnel coupling strength between lead $\ell$ and the SET dot is
 characterized by
 $\Gam_\ell(\omg)=2\pi\sum_k|t_{\ell k}|^2\dlt(\omg-\vpl_{\ell k})$.
 In what follows,  we assume wide bands in the  electrodes, which
 yields energy independent couplings $\Gam_{\rmL/\rmR}$.
 Throughout this work, we set $\hbar=e=1$
 for the Planck constant and electron charge, unless stated otherwise.

 \section{\label{thsec3}Formalism}

 \subsection{Conditional master equation}

 To achieve a description of the output from the SET detector, we
 employ the transport particle--number--resolved  reduced  density
 matrices $\rho^{(n_{\rm L},n_{\rm R})}$,  where    $n_{\rm L(R)}$
 denotes the number of electrons tunneled through the  left (right)
 junction. The corresponding \emph{conditional} quantum master equation
 reads \cite{Luo09385801,Li05205304,Luo07085325,Luo08345215,Li091707}
 \bea\label{CQME}
 \dot{\rho}^{(\nL,\nR)} &\!\!\!=\!\!\!&-\,\rmi\,{\cal L}\,
 \rho^{\,(\nL,\nR)}  - \frac{1}{2}\,\{\,[\,d^\dag\, A^{(-)}\,
 \rho^{\,(\nL,\nR)}
 \nl
 &\!\!\!\!\!\!&+\,\rho^{(\nL,\nR)} A^{(+)}d^\dag]
 \!-\![A^{(\!-\!)}_\rmL \rho^{(\nL-1,\nR)} d^\dag
 \nl
 &\!\!\!\!\!\!&+\,d^\dag\!
 \rho^{(\nL+1,\nR)}\!A^{(+)}_\rmL\!
 \!+\! A^{(-)}_\rmR\! \rho^{(\nL,\nR-1)} d^\dag
 \nl
 &\!\!\!\!\!\!&+\,d^\dag
 \rho^{(\nL,\nR+1)}A^{(+)}_\rmR]+{\rm h.c.}\},
 \eea
 where $\cal L$ is defined as
 ${\cal L}(\cdots)\equiv[H_{\rm S},(\cdots)]$, and
 $A^{(\pm)}=\sum_{\ell}A^{(\pm)}_\ell$, with
 $A^{(\pm)}_\ell\equiv[C^{(\pm)}_\ell(\pm{\cal L})+\rmi D^{(\pm)}_\ell
 (\pm{\cal L})]d$. Here $C^{(\pm)}_\ell(\pm{\cal L})=
 \int_{-\infty}^\infty dt \,C^{(\pm)}_\ell(t) e^{\pm \rmi {\cal L}t}$
 are spectral functions. The involving bath correlation functions
 are respectively $C_\ell^{(+)}(t)= \la F_\ell^\dag(t)F_\ell\ra_\rmB$,
 and $C_\ell^{(-)}(t)=\la F_\ell(t)F^\dag_\ell\ra_\rmB$, with
 $\la \cdots\ra_\rmB\equiv{\rm Tr}_{\rm B}[(\cdots)\rho_{\rm B}]$, and
 $\rho_{\rm B}$  the local  thermal equilibrium state of the  SET
 leads.
 The involved dispersion functions $D_\ell^{(\pm)}(\pm{\cal L})$
 can be evaluated via the Kramers--Kronig relation
 \be
 D_\ell^{(\pm)}(\pm{\cal L})=-\frac{1}{\pi}{\cal P}
 \int_{-\infty}^{\infty}d\omg
 \frac{C_\ell^{(\pm)}(\pm\omg)}{{\cal L}-\omg},
 \ee
 where ${\cal P}$ denotes the principal value.    Physically, the
 dispersion is responsible for the renormalization.\cite{Luo09385801,Xu029196,%
 Yan05187,Cal83587,Wei08}

 \subsection{Output current}

 With the knowledge of the  above  conditional state, the joint
 probability function for $\nL$  electrons  passed through left
 junction and $\nR$  electrons passed through right junction is
 determined  as  $P(\nL,\nR)= {\rm Tr} \rho^{(\nL,\nR)}$, where
 ${\rm Tr}(\cdots)$  denotes  the trace over the system degrees
 of freedom. The current through junction $\ell\in$\{L,R\}  then
 reads $I_\ell=\frac{d}{dt}\sum_{\nL,\nR}n_\ell P(\nL,\nR)=
 {\rm Tr}\dot{N}_\ell$, where
 $N_\ell\equiv\sum_{\nL,\nR}n_\ell P(\nL,\nR)$ can be calculated
 via its equation of motion
 \bsube
 \be\label{Nt}
 \frac{d}{dt}N_\ell=-\rmi{\cal L}N_\ell-{\cal R}N_\ell
 +{\cal T}^{(-)}_\ell \rho,
 \ee
 with
 \bea
 {\cal R}(\cdots)=-\frac{1}{2}[d^\dag,A^{(-)}(\cdots)-(\cdots)
 A^{(+)}]+{\rm h.c.},
 \\
 {\cal T}_\ell^{(\pm)}(\cdots)=\frac{1}{2}[A^{(-)}_\ell(\cdots)
 d^\dag \pm d^\dag(\cdots)A^{(+)}_\ell]+{\rm h.c.}.
 \eea
 \esube
 Straightforwardly, the transport current through junction $\ell$
 is $I_\ell(t)={\rm Tr}[{\cal T}_\ell^{(-)}\rho(t)]$.
 Here $\rho(t)$ is the \emph{unconditional} density matrix, which
 simply satisfies
 \be\label{QME}
 \dot{\rho}=-\rmi{\cal L}\rho-{\cal R}\rho.
 \ee

 \begin{figure}
 \includegraphics*[scale=0.9]{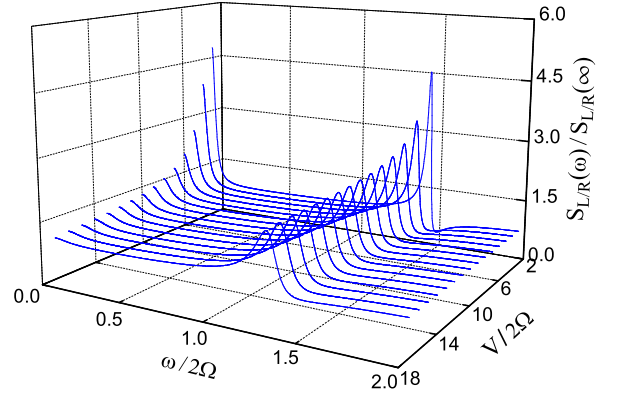}
 \caption{\label{Fig2}Noise spectral of junction currents
 scaled by its own pedestal at
 different measurement voltages for an asymmetrical SET detector
 with $\GamR/\GamL=10$ and $\tilde{E}/2\Omg=10$.}
 \end{figure}

 \subsection{Current noise spectrum}

 In continuous weak measurement of qubit oscillations,  the  most
 important output is the spectral density of the current.
 The  circuit  current  of  the  SET  detector,  according to the
 Ramo-Shockley theorem,\cite{Bla001} is
 $I(t)=\etaL I_{\rm L}+\etaR I_{\rm R}$,  where  the coefficients
 $\etaL$ and $\etaR$ depend on junction capacitances and  satisfy
 $\etaL+\etaR=1$. Together with the charge conservation
 law  $I_{\rm L}=I_{\rm R}+\dot{Q}$, where $Q$ represents the
 electron charge in the SET dot, one readily obtains
 $I(t)I(0)=\etaL I_{\rm L}(t)I_{\rm L}(0)
 +\etaR I_{\rm R}(t)I_{\rm R}(0)-\etaL\etaR\dot{Q}(t)\dot{Q}(0)$.
 Accordingly,  the circuit noise spectral is a sum of
 three parts\cite{Jia09075320,Luo07085325,Gur05205341,Agu04206601}
 \be\label{Stot}
 S(\omg)=\etaL S_{\rm L}(\omg)+ \etaR S_{\rm R}(\omg)-\etaL\etaR
 S_{\rm ch}(\omg),
 \ee
 with $S_{\rm L(R)}(\omg)$  the  noise spectral of the left (right)
 junction current, and   $S_{\rm ch}(\omg)$ the charge
 fluctuations in the SET dot.
 The  noise spectral of tunneling current $S_{\rm L/R}$ can be
 evaluated via the MacDonald's formula\cite{Mac62,Fli05411,Ela02289}
 \be\label{Salf}
 S_\ell(\omg)=2\omega\int_0^\infty\!\!dt \sin(\omg t)\frac{d}{dt}
 [\la n_\ell^{2}(t)\ra-(\bar{I}t)^2],
 \ee
 where $\bar{I}\equiv I(t\rightarrow\infty)$ is the stationary  current, and
 $\la n_\ell^2(t)\ra\equiv\sum_{\nL,\nR}n_\ell^2 P(\nL,\nR)$.
 With  the help of the conditional master equation (\ref{CQME}), it can be
 shown that
 \bea\label{n2t}
 \frac{d}{dt}\la n_\ell^2(t)\ra={\rm Tr}[2{\cal T}^{(-)}_\ell
 N_\ell(t)+{\cal T}_\ell^{(+)}\rho_{\rm st}].
 \eea
 Here $N_\ell(t)$ can be found from \Eq{Nt}, and  $\rho_{\rm st}$
 is the stationary solution of the unconditional master equation
 (\ref{QME}).

 The symmetrized spectrum for the charge fluctuation in the SET dot
 reads\cite{Luo07085325}
 \be\label{Sch}
 S_{\rm ch}(\omg)=\omg^2\int_{-\infty}^{\infty}\!d\tau\la N(\tau)N
 +NN(\tau)\ra e^{\rmi\omg\tau},
 \ee
 where $\la N(\tau)N\ra\equiv{\rm Tr}{\rm Tr}_{\rm B}[U^\dag(\tau)N
 U(\tau)N\rho_{\rm st}\rho_{\rm B}]$ with $U(\tau)$  the  evolution
 operator of the entire  system  and  $N$ the electron--number
 operator of the SET dot.
 It can be reexpressed as $\la N(\tau)N\ra={\rm Tr}[N\vrho(\tau)]$,
 where the alternative reduced density matrix
 $\vrho(\tau)\equiv{\rm Tr}_{\rm B}[U(\tau)N\rho_{\rm st}\rho_{\rm B}
 U^\dag(\tau)]$, under the standard Born approximation, satisfies the
 same equation as $\rho(t)$, but with initial condition
 $\vrho(0)=N\rho_{\rm st}$.
 Finally, the spectral of charge fluctuations is obtained
 as\cite{Luo07085325}
 \be
 S_{\rm ch}(\omg)=2\omg^2{\rm Re}\,{\rm Tr}\{N[\tilde{\vrho}(\omg) +
 \tilde{\vrho}(-\omg)]\},
 \ee
 where $\tilde{\vrho}(\omg)$ is the Fourier transform of $\vrho(t)$,   and
 satisfies
 \be
 -\rmi\omg\tilde{\vrho}(\omg)=-\rmi{\cal L}
 \tilde{\vrho}(\omg)-{\cal R}\tilde{\vrho}(\omg)+N\rho_{\rm st}.
 \ee

\section{\label{thsec4}Results and discussions}

 In this section, we will focus our analysis on the
 spectral density of the current.
 Different from the QPC detectors, the circuit current
 noise spectral of an SET detector is an appropriate
 combination of three  components, as shown in \Eq{Stot}.
 Hereafter, we first investigate the spectral of
 junction current and charge fluctuations, respectively,
 and then analyze the measurement effectiveness based on
 the circuit noise spectral.

\subsection{Spectral of junction current}

 In what follows, we assume a symmetric qubit ($\epl=0$),
 and $E=0$ for simplicity.
 In continuous weak measurement of qubit oscillation, the
 signal is manifested in the spectral density of the
 detector output as a peak at qubit oscillation frequency
 $2\Omg$.
 The spectral of junction current at different bias voltages
 is plotted in \Fig{Fig2} for an asymmetric SET detector with
 $\Gam_\rmR\gg\Gam_\rmL$.
 It is of interest to note that the peak reflecting the
 qubit oscillations shifts with the measurement voltages.
 In the high voltage regime, the oscillation peak is
 located approximately at $\omg\approx2\Omg$.
 As the measurement voltage decreases, the position
 of the oscillation peak is strongly deviated.
 Actually, this unique noise feature originates from
 the SET detection--induced
 backaction: renormalization of the qubit parameters.
 The details are explained below.

 \begin{figure}
 \includegraphics*[scale=0.77]{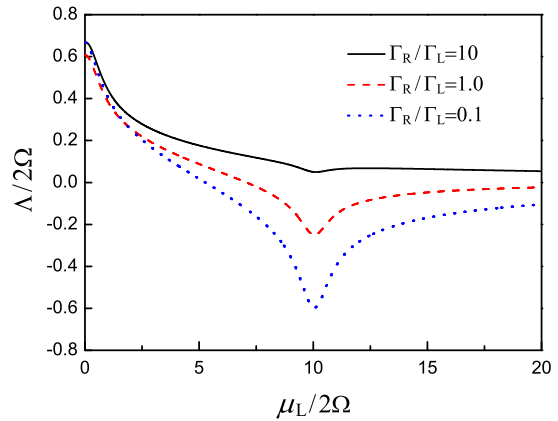}
 \caption{\label{Fig3}Qubit level renormalization versus
 measurement voltage at temperature $k_\rmB T/2\Omg=0.25$.}
 \end{figure}

 The states involved are depicted in \Fig{Fig1}.
 There are totally four eigenstates for the reduced system.
 The eigen--energies are $\lmd_+\pm\Omg$, and $\lmd_-\pm\Omg$,
 respectively, with
 $\lmd_\pm=\frac{1}{2}(\tilde{E}\pm\sqrt{\tilde{E}^2+\Omg^2})$.
 Here, we limit our discussions to the strong SET--qubit coupling
 regime ($\tilde{E}\gg\Omg$). Then the eigen--energies can be
 markedly reduced, with $\lmd_+\pm\Omg\approx\lmd_+$ and
 $\lmd_-\pm\Omg\approx\lmd_-$.
 In the bias regime $\lmd_+>\mu_\rmL>\lmd_->\mu_\rmR$, the quantum
 master equation describing the reduced dynamics can be
 greatly simplified.
 Let us denote the density matrix elements by $\rho_{jj'}$,  with
 $j,j'=$\{a,b,c,d\}. The diagonal terms of the density matrix
 $\rho_{jj}$ are the probabilities of finding the system in one
 of the electron configurations (states) as shown in \Fig{Fig1}.
 The off--diagonal matrix elements (``coherencies''), describe the
 linear superposition of these states.
 The quantum master equation in this representation simply reads
 \bsube\label{QMEele}
 \bea
 \dot{\rho}_{\rm aa}
 &=&\GamR\rho_{\rm cc}-\GamL\rho_{\rm aa}+\rmi\Omg (\rho_{\rm ab}
 -\rho_{\rm ba}),
 \\
 \dot{\rho}_{\rm bb}
 &=&(\GamL+\GamR)\rho_{\rm dd}-\rmi\Omg(\rho_{\rm ab}-\rho_{\rm ba}),
 \\
 \dot{\rho}_{\rm cc}
 &=&\GamL\rho_{\rm aa}-\GamR\rho_{\rm cc}+\rmi\Omg(\rho_{\rm cd}-\rho_{\rm dc}),
 \\
 \dot{\rho}_{\rm dd}
 &=&-(\GamL+\GamR)\rho_{\rm dd}-\rmi\Omg(\rho_{\rm cd}-\rho_{\rm dc}),
 \\
 \dot{\rho}_{\rm ab}
 &=&\rmi\Omg(\rho_{\rm aa}-\rho_{\rm bb})+\rmi\Lmd (\rho_{\rm ab}-\rho_{\rm cd})
 \nl
 && +\left(\frac{1}{2}\GamL+\GamR\right)\rho_{\rm cd}
 -\frac{1}{2}\GamL\rho_{\rm ab},\label{rhoab}
 \\
 \dot{\rho}_{\rm cd}
 &=&\rmi\Omg(\rho_{\rm cc}-\rho_{\rm dd})+\rmi\tilde{E}\rho_{\rm cd}
 -\rmi \Lmd (\rho_{\rm ab}-\rho_{\rm cd})
 \nl
 &&-\left(\frac{1}{2}\GamL+\GamR\right)\rho_{\rm cd}
 +\frac{1}{2}\GamL\rho_{\rm ab},\label{rhocd}
 \eea
 \esube
 where the involving Fermi functions in the tunneling rates are
 approximated by either one or zero. By observing the equations
 of motion of the off--diagonal matrix elements [\Eqs{rhoab} and
 (\ref{rhocd})], one finds that the qubit level detuning are
 renormalized, i.e., $\epl\rightarrow\epl+\Lmd$ with
 $\Lmd=\sum_{\ell={\rm L,R}}\{\phi_\ell(\lmd_+)-\phi_\ell(\lmd_-)\}$,
 and
 \be
 \phi_\ell(\omg)=\frac{\Gam_\ell}{2\pi}\,{\rm Re}\!\left[\Psi\left(
 \frac{1}{2}+\rmi\frac{\omg-\mu_\ell}{2\pi k_{\rmB}T}\right)\right].
 \ee
 Here $\Psi$ is the digamma function, $\mu_\ell$ is the chemical
 potential of the SET electrode $\ell\in$\{L,R\}, $k_\rmB$ the
 Boltzmann constant, and $T$ the temperature.

 In \Fig{Fig3}, the backaction--induced renormalization is
 plotted against the measurement voltage.
 The energy shift is closely related to the tunnel--coupling
 asymmetry and depends on the level positions of the reduced
 system relative to the Fermi energy.
 It is found that $\Lmd$ reaches a local
 extremum, each time when the Fermi energy of the left lead
 becomes resonant with the eigen--energies, $\mu_\rmL=\lmd_-$
 or $\mu_\rmL=\lmd_+$.
 Furthermore, the local minimum at $\lmd_+$ is markedly affected by the
 tunnel--tunneling asymmetry, i.e., the dip is more pronounced
 as the ratio $\GamR/\GamL$ decreases.

 It is right the shift of qubit level that results in
 a renormalized Rabi frequency given by
 $\omg_\rmR=\sqrt{\Lmd^2+(2\Omg)^2}$. It grows with
 decreasing measurement voltages, as implied in \Fig{Fig3}.
 Eventually, the spectral of junction current $S_{\rm L/R}(\omg)$
 exhibits the unique feature that the peak reflecting qubit
 oscillation shifts with measurement voltages,
 as shown in \Fig{Fig2}.

 \begin{figure}
 \includegraphics*[scale=0.9]{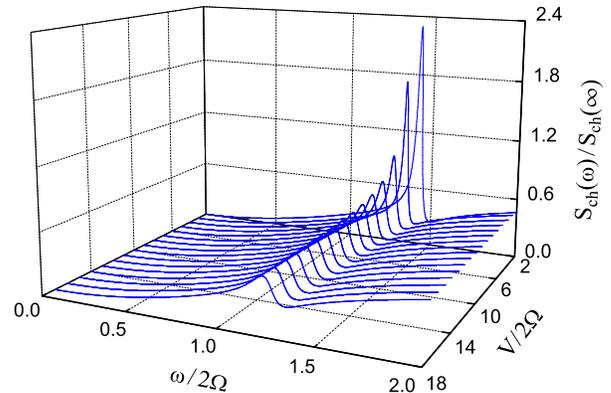}
 \caption{\label{Fig4}Spectral of charge fluctuations
 scaled by its own pedestal for different measurement
 voltages.
 The plotting parameters are the same as in \Fig{Fig2}.}
 \end{figure}

 Nevertheless, the bias--dependent peak is not the sole
 interesting behavior due to the backaction--induced
 energy shift.
 Another noticeable feature arising from the level
 renormalization is the appearance of the noise peak
 at zero frequency.
 It is known, that the peak at zero frequency is a signature of the
 quantum Zeno effect, which indicates the inhibition
 of transitions between quantum states due to continuous
 measurement.\cite{Kor01165310,Shn0211618,Gur03066801}
 The basic picture is that, due to a strong renormalized
 level shift, the detector is more readily to localize
 the electron in one of the levels for a longer time,
 leading thus to incoherent jumps between the two levels.
 Yet, the qubit coherence is not strongly destroyed.
 Eventually, the zero--frequency peak and the coherent
 peak coexist in the low bias regime, as shown in \Fig{Fig2}.
 This intriguing feature is different from that in the
 previous work,\cite{Gur05073303,Jia09075320} where
 the zero--frequency peak is not observed.

\subsection{Measurement effectiveness}

 In continuous weak measurement of quantum coherent
 oscillations of a qubit, an interesting feature of
 the output noise spectral is that
 the peak--to--pedestal (``signal--to--noise'') ratio
 provides the measure  of detector ``ideality'', i.e.,
 shows how close the detector can be quantum limited.
 For a perfectly efficient linear detector the maximum peak
 height can reach 4 times than the noise pedestal,
 which is universal and known as Korotkov--Averin
 bound.\cite{Jia09075320,Kor01165310,Kor01085312,Goa01235307,Kor01115403}
 Physically, this limit arises from the tendency of
 quantum  measurement to localize the system in one
 of the measured eigenstates.
 So far, many schemes such as quantum nondemolition
 measurement,\cite{Ave02207901,Jor05125333} quantum
 feedback control,\cite{Wan07155304} and measurement
 by using two detectors\cite{Jor05220401} have been
 proposed to overcome this bound.

 Different from that of a QPC detector, the circuit
 noise spectral of an SET detector, however, consists
 of three parts as shown in \Eq{Stot}.
 It is therefore of vital importance to investigate
 the charge fluctuations in the SET dot, which uniquely
 characterizes the current fluctuations in SET dot, and
 is intimately associated with the qubit dynamics.
 The numerical result of the charge fluctuations at
 different measurement voltages is displayed in \Fig{Fig4}.
 At low frequencies the charge fluctuations are strongly
 suppressed.
 The most prominent signal is the qubit oscillation
 peak which is located at the renormalized Rabi
 frequency $\omg_\rmR$, and varies with measurement
 voltages.
 It is important to note that the charge fluctuations
 has an essential role to play in the signal--to--noise
 ratio, which will be revealed later.

\begin{figure}
 \includegraphics*[scale=0.7]{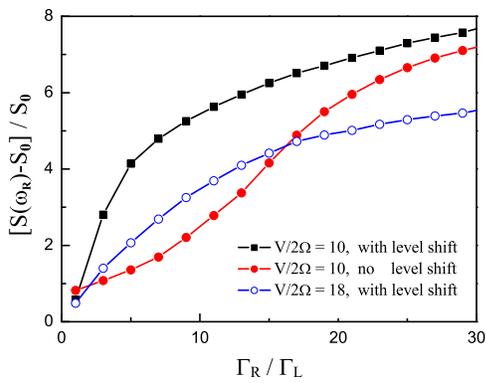}
 \caption{\label{Fig5}''Signal--to--noise'' ratio
 versus the degree of tunnel--coupling asymmetry $\Gam_\rmR/\Gam_\rmL$
 for different measurement voltages.
 Other plotting parameters are the same as in \Fig{Fig2}.}
 \end{figure}

 With the knowledge of the junction current noise and
 the spectral of charge fluctuations, the circuit noise
 can be readily obtained. It allows us to evaluate the
 signal--to--noise ratio
 \bea
 {\rm SNR}=\frac{S(\omg_\rmR)-S_0}{S_0},
 \eea
 where $S_0\equiv S(\omg\rightarrow\infty)$ is the
 pedestal of circuit noise.
 The numerical result of SNR versus tunnel--coupling
 asymmetry $\Gam_\rmR/\Gam_\rmL$ is plotted in \Fig{Fig5},
 where we have assumed symmetric capacitive--coupling
 ($\eta_\rmL=\eta_\rmR=\frac{1}{2}$). In this case,
 the influence of charge fluctuations on the circuit
 noise is maximized.

 A remarkable feature observed is that the SNR
 can exceed ``4'', i.e., the upper bound for any
 linear--response detectors.\cite{Kor01165310}
 This seems to be at variance with
 Ref. \onlinecite{Gur05073303}, which predict
 that the effectiveness of an asymmetric SET
 detector can not reach that of an ideal detector.
 Here the large spectral of charge fluctuations is
 responsible for the violation of the Korotkov-Averin
 bound, as its pedestal can markedly reduces that of
 the circuit noise [cf. \Eq{Stot}].
 The SNR is thereby enhanced, and finally exceeds the
 upper  bound ``4''.
 Recently, a measurement scheme of a charge qubit
 with two quantum point contacts was proposed by
 Jordan and B\"{u}ttiker.\cite{Jor05220401}
 They also observed the violation of the Korotkov--Averin
 bound, which is ascribed to the negligible small spectral
 of junction cross--correlations.\cite{Jia09075320,Jor05220401}
 By heuristically viewing the left and right SET
 electrodes as separate detectors, their analysis
 qualitatively support our results.

 To clearly demonstrate the effect of level renormalization,
 in \Fig{Fig5} we have also displayed the result in the
 absence of energy shift. It is found that the renormalization
 can strongly enhance the SNR.
 Particularly, in the regime of low $\GamR/\GamL$ ratio, the
 effectiveness exceeds the Korotkov--Averin bound purely due
 to the energy renormalization.
 It suggests that an ideal SET detector can be achieved
 even though the tunnel--coupling is not strongly asymmetric.
 This finding is different from that in Ref.~\onlinecite{Jia09075320},
 where a very large tunnel--coupling asymmetry is required
 to overcome the Korotkov--Averin bound.

 Finally, let us accentuate the influence of the measurement
 voltage on the signal--to--noise ratio.
 At a relative large voltage (e.g. $V/2\Omg=18$), the state
 (d) shown in \Fig{Fig1} becomes partially available at
 finite temperature.
 The dynamics of the system is described by the corresponding
 master equation, which is similar to \Eq{QMEele}, but with the
 crucial difference of an increased decoherence rate.
 Yet, such an increasing of the decoherence rate is
 associated with the detector shot noise, rather than
 the information
 flow.\cite{Cle03165324,Gur05073303,Kor01085312,Kor01115403}
 Eventually, as shown in \Fig{Fig5}, the signal--to--noise
 ratio is reduced in comparison with that in the lower
 voltage situation.

\section{\label{thsec5}Summary}

 In summary, the problem of qubit measurements by a
 single electron transistor detector is investigated,
 with special attention being paid to the renormalization
 effect.
 Our analysis reveals that the dynamical renormalization,
 which was neglected in previous studies, can strikingly
 influence the spectral of the detector output.
 It is therefore of essential importance to take this
 effect into consideration for correct analyzing and
 understanding the measurement results.
 Under proper tunnel--coupling asymmetry, the effectiveness
 of the single electron transistor detector can be considerably
 increased.
 Remarkably, it is observed that the signal--to--noise
 ratio can  exceed the universal
 Korotkov--Averin bound due purely to the dynamical
 renormalization.

 \vspace{1cm}
 \noindent{\bf Acknowledgements}\\

 Support from the National Natural Science Foundation of
 China (Grants Nos. 10904128 and 11004124) is
 gratefully acknowledged.



\end{document}